\documentclass[preprint,12pt]{elsarticle}

\usepackage{booktabs}
\usepackage{amsmath}
\usepackage{verbatim}
\usepackage{enumerate}
\usepackage{siunitx}
\usepackage[shortlabels]{enumitem}
\usepackage{fancybox}
\usepackage[hyphens]{url}
\usepackage{hyperref}
\usepackage{flushend}
\usepackage{graphicx}
\usepackage{setspace}
\usepackage[roman]{parnotes}
\usepackage{color}
\usepackage{colortbl}
\usepackage{listings}
\usepackage{tabulary}
\usepackage{xcolor}
\usepackage{framed}
\usepackage{subfigure}
\usepackage{xspace}
\usepackage{makecell}
\usepackage[symbol]{footmisc}
\usepackage{balance}
\usepackage{soul}
\usepackage{csquotes}
\usepackage{dirtytalk}

\usepackage{adjustbox}
\usepackage{multirow}

\usepackage{xspace}
\newcolumntype{L}{>{\arraybackslash}m{8cm}}
\newcolumntype{T}{>{\arraybackslash}m{13cm}}
\newcolumntype{C}[1]{>{\centering\let\newline\\arraybackslash\hspace{0pt}}m{#1}}
\newcolumntype{R}[1]{>{\raggedleft\let\newline\\arraybackslash\hspace{0pt}}m{#1}}

\newcounter{observation}
\newcommand{\observation}[1]{\refstepcounter{observation}
	\begin{center}
		\framebox{
			\begin{minipage}{0.93\columnwidth}
				{} \textit{#1}
			\end{minipage}
		}
	\end{center}
}

\makeatother

\setlist{nosep}
\setlist[itemize]{leftmargin=2.3em}
\setlist[enumerate]{leftmargin=2.3em}
\journal{Information and Software Technology}

\begin{document}

\begin{frontmatter}



\title{Using AI-Based Coding Assistants in Practice:\\State of Affairs, Perceptions, and Ways Forward}






\author[JB]{Agnia Sergeyuk\fnref{contrib}}
\fntext[contrib]{The first two authors contributed equally to this work}
\ead{agnia.sergeyuk@jetbrains.com}

\author[JB]{Yaroslav Golubev\corref{mycorrespondingauthor}\fnref{contrib}}
\cortext[mycorrespondingauthor]{Corresponding author}
\ead{yaroslav.golubev@jetbrains.com}

\author[JB2]{Timofey Bryksin}
\ead{timofey.bryksin@jetbrains.com}

\author[UCI]{Iftekhar Ahmed}
\ead{iftekha@uci.edu}

\address[JB]{JetBrains Research, Belgrade, Serbia}
\address[JB2]{JetBrains Research, Limassol, Cyprus}
\address[UCI]{University of California, Irvine, CA, United States}

\begin{abstract}

\textbf{Context}.
The last several years saw the emergence of \textit{AI assistants} for code --- multi-purpose AI-based helpers in software engineering. As they become omnipresent in all aspects of software development, it becomes critical to understand their usage patterns.

\textbf{Objective}.
We aim to better understand \textit{how specifically} developers are using AI assistants, why they are \textit{not} using them in certain parts of their development workflow, and what needs to be improved in the future.

\textbf{Methods}.
In this work, we carried out a large-scale survey aimed at how AI assistants are used, focusing on specific software development activities and stages. We collected opinions of 481 programmers on five broad activities: (a) implementing new features, (b) writing tests, (c) bug triaging, (d) refactoring, and (e) writing natural-language artifacts, as well as their individual stages.

\textbf{Results}.
Our results provide a novel comparison of different stages where AI assistants are used that is both comprehensive and detailed. It highlights specific activities that developers find less enjoyable and want to delegate to an AI assistant, \textit{e.g.}, writing tests and natural-language artifacts. We also determine more granular stages where AI assistants are used, such as generating tests and generating docstrings, as well as less studied parts of the workflow, such as generating test data. Among the reasons for not using assistants, there are general aspects like trust and company policies, as well as more concrete issues like the lack of project-size context, which can be the focus of the future research.

\textbf{Conclusion}.
The provided analysis highlights stages of software development that developers want to delegate and that are already popular for using AI assistants, which can be a good focus for features aimed to help developers right now. The main reasons for not using AI assistants can serve as a guideline for future work.

\end{abstract}


\end{frontmatter}

\section{Introduction}
\label{sec:introduction}

In the ever-evolving landscape of technology, the symbiotic relationship between humans and Artificial Intelligence (AI) has ushered in a new era of innovation, transcending traditional boundaries and catalyzing unprecedented advancements. The recent advent of Large Language Models (LLMs) is one such advancement that, like many other fields, is significantly impacting software development~\cite{kaddour2023challenges}.

LLM-powered AI assistants, such as GitHub Copilot~\cite{githubcopilot}, JetBrains AI Assistant~\cite{jetbrainsai}, and Visual Studio IntelliCode~\cite{visualstudiointellicode}, have emerged as invaluable assets. They assist developers in code generation~\cite{openaicodex}, bug fixing~\cite{berabi2024deepcode}, refactoring~\cite{codescene}, testing~\cite{testsigma}, and almost every aspect of the software development life-cycle. Being trained on vast corpora of data, these assistants provide intelligent suggestions and recommendations, thereby augmenting the capabilities of software developers and rapidly gaining popularity among them~\cite{stackoverflowsurvey}.

Given their rapid growth, understanding software developers' perceptions and needs regarding AI assistants is paramount in shaping the future of software development. Recent studies have started looking into why AI assistants are used, what their high-level drawbacks are~\cite{wang2023practitioners, zhou2023concerns, liang2024large}, etc. However, the existing studies do not consider individual activities within the software development life-cycle and their specificities. Because of this, it is challenging to transform the results of such works into actionable implications, since they only deal with general concerns like quality and security, without taking into account their concrete manifestations in the developer's workflow. Conversely, other works focus on individual narrow stages, \textit{e.g.}, code completion, test generation, etc., and study them in detail. While they are more specific, it is difficult to compare and prioritize their results, because they do not take the broader picture into account. The quick development of AI assistants requires a comprehensive yet granular analysis that considers the full software development lifecycle and its detailed steps at the same time, thus allowing to find areas where AI assistants are already employed to focus on them in the short term and identify shortcomings to address them in the future. To the best of our knowledge, no study has yet been conducted on such an analysis.

To overcome the existing research gap, we carried out a large-scale survey to determine where developers need AI support and what kind of support they need. More specifically, we developed, piloted, and ran a survey that focused on answering the following three research questions:

\begin{enumerate}[font={\bfseries},label={RQ\arabic*},leftmargin=1.2cm]
    \item What is the general usage and perception of AI assistants among programmers?
    \item To what extent do programmers utilize AI assistance for specific activities in software development, and how do they perceive these activities?
    \item What are the reasons for not using AI assistants, and therefore, what needs to be improved?
\end{enumerate}

In particular, we studied the following five broad SE activities: \textbf{(a)} implementing new features, \textbf{(b)} writing tests, \textbf{(c)} bug triaging, \textbf{(d)} refactoring, and \textbf{(e)} writing natural-language artifacts, as well as their individual stages. Our survey included 38 main questions, of them 5 being open-ended, which allowed the participants to freely express their thoughts. The survey attracted 547 complete responses, and after careful filtering, we study 481 responses. Our sample is experienced and diverse, with almost half of the respondents having more than 10 years of professional experience covering all the major programming languages and types of developed software.

Overall, the contributions of this work are the following:

\begin{itemize}
    \item \textbf{Usage patterns}. We find that 84.2\% of the respondents occasionally or regularly use AI assistants, that \textit{Implementing new features} is the most popular activity to use assistants, and that among individual stages, generating and summarizing code are the most widely used. 

    \item \textbf{Areas of focus}. Taking into account which activities the developers find less enjoyable and want to delegate, as well as what stages AI assistants are used on, we highlight areas where the research needs to focus on to bring value to users right now. This includes \textit{Writing tests} in general and \textit{generating data and resources for tests} in particular, as well as \textit{Writing natural language artifacts}.

    \item \textbf{Reasons for not using AI}. We provide 20 distinct themes that represent different reasons why developers are not using AI assistants and report the main ones for all activities. The most popular ones are \textit{Lack of need for AI assistance}, \textit{AI-generated output being inaccurate}, \textit{User's lack of trust and the desire to feel in control}, and \textit{Lack of understanding context by AI assistant}.

    \item \textbf{Areas of future improvement}. Based on the prevalence of different reasons, we formulate the main areas of future work that are needed to overcome the shortcomings that the respondents describe. This includes the improvement of base systems, better integration into developers' workflow, and more actively educating and informing users about the assistants' capabilities and limitations.

\end{itemize}

The detailed description of the survey, the collected responses, as well as some additional results that did not fully fit into the paper can be found in supplementary materials~\cite{artifacts}.

The remainder of this paper is organized as follows. In Section~\ref{sec:rw}, we discuss the related work. In Section~\ref{sec:methodology}, we describe how we developed and piloted our survey, how we found the participants, and how we analyzed the results. Section~\ref{sec:results} presents these results in detail, and Section~\ref{sec:discussion} describes their specific implications. Finally, we describe the potential threats to the validity of our study in Section~\ref{sec:ttv} and conclude the paper in Section~\ref{sec:conclusion}.
\section{Related Work}
\label{sec:rw}

AI assistance is a very active area of research, and numerous studies have investigated its impact on how programmers code. In this section, we describe the main surveys and studies --- both academic and industrial --- that explore the users' perceptions and benefits from AI assistants.

Liang et al.~\cite{liang2024large} conducted an exploratory qualitative study on the usage of AI programming assistants, highlighting motivations for usage, usability challenges, and implications for creators and users of these tools. The study had 410 developers as participants. 
The authors concluded that while using AI to reduce keystrokes, finish programming tasks quickly, and recall syntax, developers sometimes struggle to receive AI outputs that align with their requirements and expectations.

Wang et al.~\cite{how2023wang} present a study that aims to understand practitioners' expectations on code completion, compare them with existing research, and highlight the need for researchers to develop techniques that meet practitioners' demands. The methodology involves semi-structured interviews with 15 professionals and an exploratory survey with 599 professionals.
The authors found that practitioners expect code completion tools to work for different granularities and scenarios. They also expect a tool to be accurate, display personalized completion, be available offline, and be relatively fast.

Ziegler et al.~\cite{Ziegler2022Productivity} discuss the use of neural code synthesis in software development and study perceived productivity by investigating whether usage measurements of developer interactions with GitHub Copilot can predict perceived productivity as reported by developers in a survey. 
The authors found that the acceptance rate of shown suggestions is a better predictor of perceived productivity than more specific metrics regarding the persistence of completions in the code over time. This suggests that the rate at which suggestions are accepted drives developers' perception of productivity. 

Zhou et al.~\cite{zhou2023concerns} aimed at understanding the issues and challenges developers face when using Copilot in practice, as well as their underlying causes and potential solutions. The results of the analysis of data from 476 GitHub issues, 706 GitHub discussions, and 142 Stack Overflow posts highlight that users commonly face Operation and Compatibility Issues with Copilot. The most frequent causes include Copilot Internal Issues, Network Connection Problems, and Editor/IDE Compatibility Issues. Predominant solutions involve Bug Fixes by Copilot, Configuration/Setting Modifications, and Using Suitable Versions. The authors suggest integrating a code explanation feature to enhance Copilot's overall utility and effectiveness in practical development scenarios, considering the additional time required for code suggestion verification.

Pothukuchi et al.~\cite{pothukuchi2023impact} examine the potential of generative AI to automate and enhance traditional software development practices, thereby improving efficiency and reducing costs. A total of 30 professionals participated in the interview study and highlighted significant improvements in development speed and code quality. The authors indicate that generative AI can significantly transform the SDLC by automating repetitive tasks and providing intelligent code suggestions, and propose a new model called Generative AI-Assisted Software Development Lifecycle.

Mozannar et al.~\cite{mozannar2023reading} conducted a comprehensive study to inform the understanding of how developers interact with AI tools and how to improve this experience. They studied the impact of GitHub Copilot on programmers' behavior during coding sessions. As a result, they identified 12 common programmer activities related to AI code completion systems. The researchers found that developers spend more time reviewing code than writing it. 

Barke et al.~\cite{Barke2023Grounded} propose a grounded theory of the bimodal nature of programmers’ interactions with AI assistance that states that there are two main interaction modes between which developers fluidly switch while programming --- exploration and acceleration. In acceleration mode, the programmer already knows what they want to do next, and AI helps them get there quicker; interactions in this mode are fast and do not break programmer’s flow. In exploration mode, the programmer is not sure how to proceed and uses Copilot to explore their options or get a starting point for the solution; interactions in this mode are slow and deliberate, and include explicit prompting and more extensive validation.

Moreover, several companies from industry conducted large-scale studies to form an understanding of the market of AI tooling for coding. 
A survey of 89,184 developers conducted in 2023 by Stack Overflow~\cite{stackoverflowsurvey} found that 70\% are using or planning to use AI tools, with 77\% expressing a favorable view. Respondents noted increased productivity with these tools, as well as their trust in their accuracy. The most recent Stack Overflow survey~\cite{stackoverflowsurvey2024}, conducted on 1700 people, revealed the continuation of the trend --- users report more quality work time.

A GitHub survey~\cite{githubblog} of 500 non-manager developers found that 67\% have used AI tools both at work and in personal projects. 70\% believe AI coding tools will benefit their work, primarily for upskilling and productivity gains. 81\% expect these tools to enhance team collaboration, particularly in security reviews, planning, and pair programming.

The JetBrains Developer Ecosystem Survey~\cite{jetbrainsdevsurvey} found that developers are optimistic about AI advancements and actively use its capabilities despite security and ethical concerns. 59\% have security concerns, 42\% ethical concerns, yet 53\% are willing to use generative AI services for work. Most commonly, developers use AI to ask general software development questions and generate code, comments, or documentation.

In general, existing research on the utilization of AI assistants covers a range of perspectives, with some studies focusing on a single activity, such as code completion, while others offer more generalized insights that may overlook the nuances of different specific activities within software development. In contrast to this existing body of work, our study aims to combine the strengths of both approaches to allow for a more direct comparison of various areas of application of AI assistants. Taking into account the entire software development lifecycle, while also asking about specific low-level work aspects, can allow our findings to be both concrete and comprehensive. Having a wide and detailed palette of opinions within one large-scale survey provides an opportunity to prioritize both short-term improvements to features that are already in use and long-term research into reasons why some features are not yet widely employed.

\section{Methodology}
\label{sec:methodology}

Our survey was aimed at answering the following three research questions:

\begin{enumerate}[font={\bfseries},label={RQ\arabic*},leftmargin=1.2cm]
    \item What is the general usage and perception of AI assistants among programmers?
    \item To what extent do programmers utilize AI assistance for specific activities in software development, and how do they perceive these activities?
    \item What are the reasons for not using AI assistants, and therefore, what needs to be improved?
\end{enumerate}

In this section, we describe our survey, the way we attracted our participants, and the analysis we carried out on the gathered data.

\subsection{Survey Structure}

The full survey, including all the stages of activities, is available in our supplementary materials~\cite{artifacts}.

Firstly, we ask the developers an array of general demographic questions (years of professional experience, employment status) and questions about their jobs (primary programming languages, job role, job level, and type of developed software). Next, to answer \textbf{RQ1} and establish the general usage and perception of AI assistants, we inquire about the participants’ familiarity with specific popular AI assistant tools, as well as their opinions about the main qualities of the code generated by AI assistants: usability, accuracy, alignment with non-functional requirements, and security.

The remaining survey is divided into five parts that correspond to five large divisions of software activity in research and in practice~\cite{lehman1980programs}, which includes \textbf{(1)} implementing new features, \textbf{(2)} writing tests, \textbf{(3)} bug triaging, \textbf{(4)} refactoring, and \textbf{(5)} writing natural language artifacts. Within each of these blocks, the questions follow the same pattern. 

First, to answer \textbf{RQ2} and discover how participants perceive the activities, we ask them to rate each activity on a 5-point Likert scale in three dimensions:

\begin{enumerate}[1.]
    \item From \textit{Unpleasant} to \textit{Enjoyable}.
    \item From \textit{Difficult} to \textit{Easy}.
    \item From \textit{Would do it myself} to \textit{Would delegate to an AI assistant}.
\end{enumerate}

Then, we deepen our understanding of activities by considering their individual \textit{stages}, which allows us to make the results of this research not only comprehensive but also detailed. For each activity, we provide a list of \textit{stages}, or component steps, within this activity and ask the participant whether they employ an AI assistant on this stage. Some stages are the same for all activities, for example, \textit{``Chatting with an AI assistant to brainstorm ideas''}, while the majority are unique, like \textit{``Generating data and resources for tests (e.g., inputs and outputs)''} for writing tests, or \textit{``Generating commit messages''} for writing natural-language artifacts.  Also, the participants could enter their own steps in the \textit{``Other''} field.

Finally, to answer \textbf{RQ3}, for each activity, we ask an open-ended question about what prevents the participants from using AI assistants in the stages they did not select and whether they would use an AI assistant in those steps if the issues were resolved. Responses to these questions can allow us to not only quantitatively determine the least popular stages, but also establish distinct reasons for not using assistants there, thus leading the way to fixing them.

To ensure the robustness of our survey, we consulted experts in survey design and carried out five pilot runs with our colleagues. The experts included members of the Surveys Team and the Research Core team at JetBrains, who specialize in designing surveys and other forms of quantitative and qualitative research. The list of pilot participants was also diverse --- two software developers, two ML engineers, and one UX researcher. Overall, the feedback from the experts and the pilot participants was used to update and refine the survey. In particular, we compiled and polished the list of stages for each activity. Also, the participants highlighted the need to have \textit{``Chatting with an AI assistant to brainstorm''} as an option for each activity, as well as having an explicit \textit{``None''} option.

\subsection{Data Collection}

The participants were invited from both Industry and Academia communities. The survey link was advertised via personal social media by the authors of the paper, three of whom are the employees of JetBrains Research, and was emailed to the list of people curated by JetBrains who gave their consent for contact with research purposes. Importantly, not all people in this mailing list are customers of JetBrains products, and the list includes the users of various IDEs and AI assistants. As a thank you, participants were given the opportunity to enter a draw for one of five 50 USD Amazon eGift Cards or an equivalent-value JetBrains product pack. The study was conducted in line with the company’s ethical standards, adhering to the values and guidelines outlined in the ICC/ESOMAR International Code~\cite{iccesomar}.

\subsection{Data Analysis}

After collecting the responses, we analyzed them. This includes filtering the results, calculating correlations, and processing open-ended questions. 

\textbf{Filtering results.} At the end of the three weeks during which the survey remained open, we received 780 responses, of which 547 were complete. Among these 547 responses, 542 came from the mailing list, indicating a strong industrial backing of our sample. This allowed us to collect diverse opinions from experienced developers acquainted with a variety of different tools, as evidenced further in Sections~\ref{sec:demographics} and \ref{sec:results:rq1}.

To ensure the quality of our data, we further applied two checks. Firstly, research shows that responses that took too little or too long time to complete are less reliable~\cite{meade2012identifying}. To filter them out, we followed an established practice of filtering out outliers, applying Tukey's fences~\cite{Wobbe2007outlier}. Because our survey was designed to facilitate quick responses and not fatigue the participant, most of the responses are relatively quick (the median time of all complete responses is 11 minutes), so this filtering did not affect quick responses. However, it did filter out 66 responses that took 30 or more minutes to complete, with some of them taking several hours. Additionally, we checked our results for responses where the first option is selected for each question~\cite{meade2012identifying}. We did not find such cases. Thus, the final data consisted of 481 responses. 

\textbf{Calculating correlations (RQ2).} For each activity, we calculate correlations for Likert-scale values of the enjoyability, easiness, and delegatability of the activity. For all correlations, we used the Spearman correlation coefficient, which is recommended to be used for Likert scales~\cite{murray2013likert}.

\textbf{Processing open-ended questions (RQ3).} For questions about why the respondents do not use AI assistants, we used thematic analysis~\cite{fereday2006demonstrating} to group the responses. The first stage of the analysis was \textit{open-coding the responses}. We decided to process all the responses together, going activity by activity: firstly, \textit{implementing new features}, then \textit{writing tests}, etc. This way, the obtained codes and later themes would be general enough, however, activity-specific codes would not be lost. The first two authors independently read all the responses and independently coded them. 
Then, the two authors discussed their codes until they reached a consensus for each response using a negotiated agreement approach~\cite{forman2007qualitative}.

From 481 respondents in five open-ended questions, we got 1,764 responses. After filtering out responses such as \textit{``NA''}, \textit{``Nothing''}, \textit{``Don't know''}, responses that were written in languages other than English, and responses that did not provide any clear criticism or reason, 1,559 responses were left and resulted in 147 different codes.

The second stage of our analysis was \textit{merging the codes into themes}. This task was performed independently by the same first two authors. Following the established methodology~\cite{braun2006using}, both authors iteratively merged similar codes into themes. Then, the authors discussed their themes and the position of each code within them until they reached a consensus. 
This process resulted in a total of 20 distinct themes. You can find the list of themes and their corresponding codes in our supplementary materials~\cite{artifacts}.
\section{Results}
\label{sec:results}
\subsection{Demographics}
\label{sec:demographics}
Before moving to the results themselves, let us briefly showcase the demographics of our studied sample. An extended list of the answers to all the demographic questions can be found in supplementary materials~\cite{artifacts}. 

The respondents came from 71 different countries and territories from all the continents. They use more than 30 diverse languages, with the most popular being \textit{Python} (45.5\% of respondents), \textit{JavaScript} (31.6\%), and \textit{Java} (26.4\%). They also develop diverse types of software, including \textit{Websites} (51.1\%), \textit{Utilities} (36.4\%), and \textit{Libraries / Frameworks} (30.6\%).

Our sample consists mostly of experienced programmers, with 48.9\% having more than 10 years of experience and 28.9\% having more than 15. 74.4\% of our sample are fully employed in corporations and organizations, with some more being partially employed, self-employed, or freelancing. The majority (86.7\%) identify themselves as \textit{Software developers}, but the sample also contains some of the other major technical positions: \textit{DevOps engineers}, \textit{Architects}, \textit{Team leads}, etc.

\subsection{RQ1: General Usage and Perception of AI Assistants in SE}
\label{sec:results:rq1}

Our first RQ focused on the current state and perceptions of using AI assistants in software development.

\begin{figure}[t]
  \centering
  \includegraphics[width = 0.8\columnwidth]
  {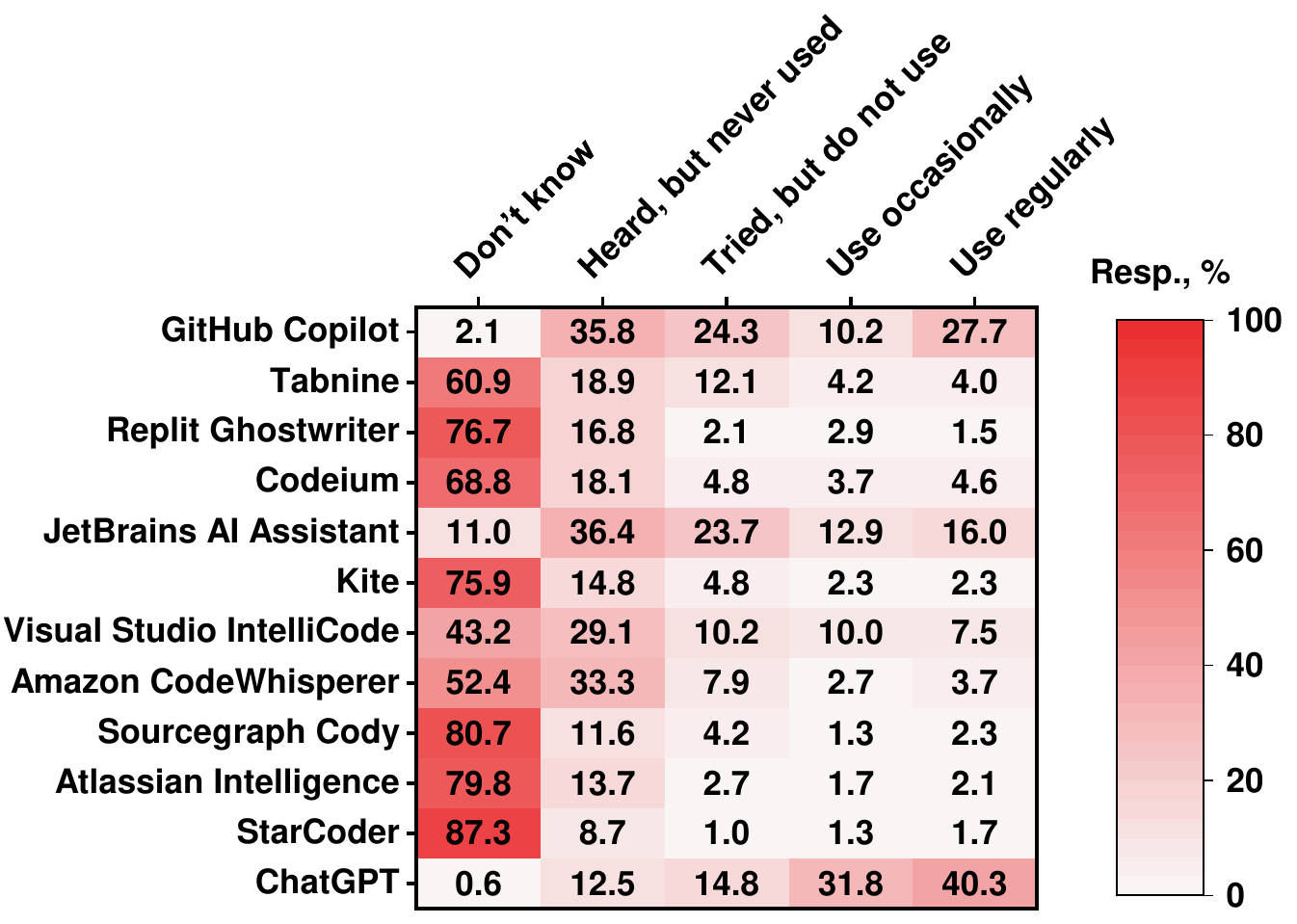}
  \caption{The tools used by the respondents.}
  \label{fig:tools}
\end{figure}

\textbf{Used tools.} The heatmap in Figure~\ref{fig:tools} shows specific tools the respondents use in their work. Overall, 84.2\% of the respondents mentioned that they use at least \textit{some} tool occasionally or regularly. In terms of specifics, we can see \textit{ChatGPT} (72.1\%), \textit{GitHub Copilot} (37.9\%), \textit{JetBrains AI Assistant} (28.9\%), and \textit{Visual Studio IntelliCode} (17.5\%) are the most frequently used tools. As for the other tools, more than half of the respondents do not know about them. This demonstrates that the field of AI assistants is dominated by just several big players and smaller ones are used more rarely. 
In the write-in responses, the most popular one was \textit{Google Gemini/Bard} (6.4\% of respondents), with some others mentioned being \textit{Ollama} and \textit{Perplexity}.

\begin{figure}[t]
  \centering
  \includegraphics[width=0.8\columnwidth]{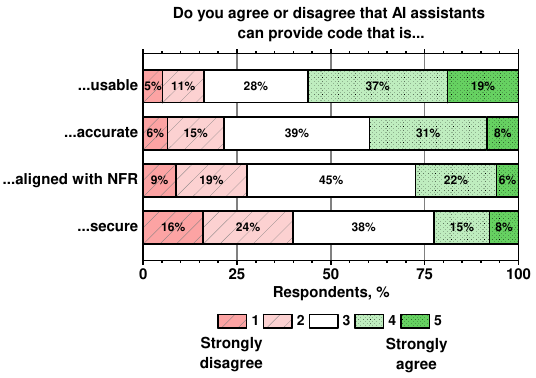}
  \caption{General opinions of respondents about code provided by AI assistants. NFR stands for non-functional requirements.}
  \label{fig:general}
\end{figure}

\textbf{Opinions about AI-generated code.} Next, Figure~\ref{fig:general} demonstrates the general opinions of the respondents about the code provided by AI assistants. We can see that the quality most positively rated is \textit{usable}, with 56\% of respondents agreeing or strongly agreeing with it. 
39\% of respondents agree or strongly agree that the code provided by AI is \textit{accurate}, while 39\% neither agree nor disagree with this statement. Regarding the \textit{alignment with non-functional requirements}, 28\% of respondents are positive, and 45\% are neutral. Finally, the most negative opinion is shared about the code being \textit{secure}, with just 23\% of respondents agreeing and as many as 40\% of them disagreeing to some extent.

We also separately checked these opinions from the part of our sample who do not use AI assistants (15.8\% of respondents who did not select using any tool occasionally or regularly). While they are a minority, their opinion is much more negative: even for the most positive quality of \textit{accurate}, only 21\% of them agree, and 45\% disagree, whereas for the other three qualities, more than half disagree. This might indicate that increasing the adoption of AI assistants requires overcoming the corresponding issues and convincing developers.

\observation{\textbf{Takeaway 1}: The overall opinion regarding code provided by AI assistants differs, with the most positive aspect being its usability and the most negative being its security.}

\subsection{RQ2: Activities and Their Stages}
\label{sec:results:rq2}

Our second RQ focused on how exactly AI assistants are used in specific SE activities and specific stages within them.

\begin{figure}[t]
  \centering
  \includegraphics[width=\textwidth]{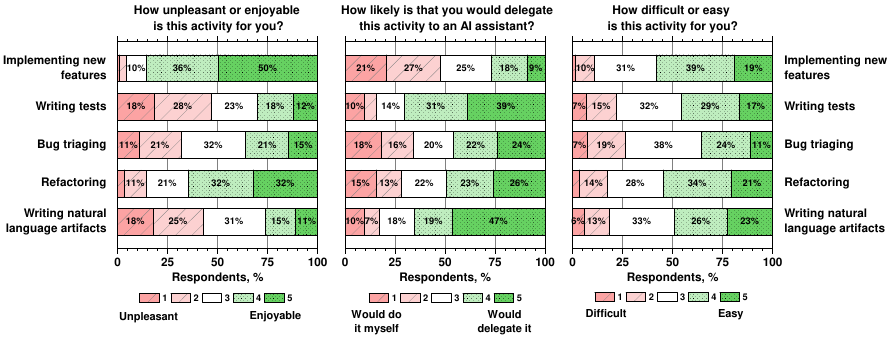}
  \caption{The general opinions of respondents about each of the five main activities.}
  \label{fig:dichotomies}
\end{figure}

\textbf{Activities.} Firstly, we compile and compare the more general opinions about the activities themselves to highlight the context of the importance of AI assistance in those activities. For each of them, we asked the participants \textbf{(1)} how unpleasant or enjoyable it is, \textbf{(2)} how difficult or easy it is, \textbf{(3)} how likely it is that they would delegate it to the AI assistant, in the form of Likert scales. The results are presented in Figure~\ref{fig:dichotomies}. We can see some noteworthy differences between the activities.

\textit{Implementing new features} is the most enjoyable activity, with as many as 86\% of respondents rating it positively. It is also the least likely activity to be delegated to an AI assistant, with 48\% of respondents answering negatively.
On the other hand, \textit{writing tests} and \textit{writing natural language artifacts} are both noticeably the least enjoyable of activities (46\% negative scores for tests and 43\% negative scores for artifacts), and they are the ones that the developers want to delegate the most --- 70\% and 66\%, respectively.

\textit{Refactoring} is the second most enjoyable activity, with 64\% positive assessment. Nonetheless, 49\% of the participants would delegate that activity to an AI assistant.
People hold more mixed opinions regarding \textit{bug triaging} --- 32\% of respondents marked it as unpleasant and 36\% as enjoyable, with 34\% unwilling and 46\% willing to delegate it to AI.

\begin{table}[t]
\small
    \centering
    \caption{Spearman correlations between answers to questions about \textbf{Enj}oyability, \textbf{Eas}iness, and \textbf{Del}egability of each of the five main activities. The underlined values are statistically significant with $p < 0.01$.}
    \vspace{0.2cm}
    \begin{tabular}{c c c}
    \toprule
    \textbf{Activity} & \textbf{Enj. / Del.} & \textbf{Eas. / Del.}\\
    \midrule
    New features & -0.04 & \underline{0.12} \\
    Writing tests & \underline{-0.24} & -0.08 \\
    Bug triaging & \underline{-0.20} & -0.08 \\
    Refactoring & \underline{-0.14} & 0.09 \\
    NL artifacts & \underline{-0.32} & 0.01 \\
    \bottomrule
    \end{tabular}
    \label{tab:correlations}
\end{table}

We performed a Spearman correlation~\cite{murray2013likert} to examine the relationship between enjoyment and willingness to delegate the activity. Presented in the first column of Table~\ref{tab:correlations}, the results reveal a negative correlation between these aspects, indicating that if a person enjoys the given activity less, they are more likely to delegate it. Notably, writing natural artifacts shows a moderately negative correlation ($r = -.32, p < .01$). 

Additionally, we explored the correlation between the difficulty of the activity and the willingness to delegate it. Overall, from Figure~\ref{fig:general}, it can be seen that opinions about the easiness of activities are more uniform, without as big differences as for other qualities. The results, presented in Table~\ref{tab:correlations} support the idea that the correlation between the difficulty of the activity and willingness to delegate is weak and mostly not statistically significant, meaning that there is no connection between these two aspects of activity for our sample. 

\observation{\textbf{Takeaway 2}. Among the main activities studied, implementing new features is the most enjoyable and the least likely to be delegated to an assistant, while writing tests and writing natural-language artifacts are the most unpleasant and the most likely to be delegated.}

\begin{figure}
  \centering
  \includegraphics[width=\textwidth]{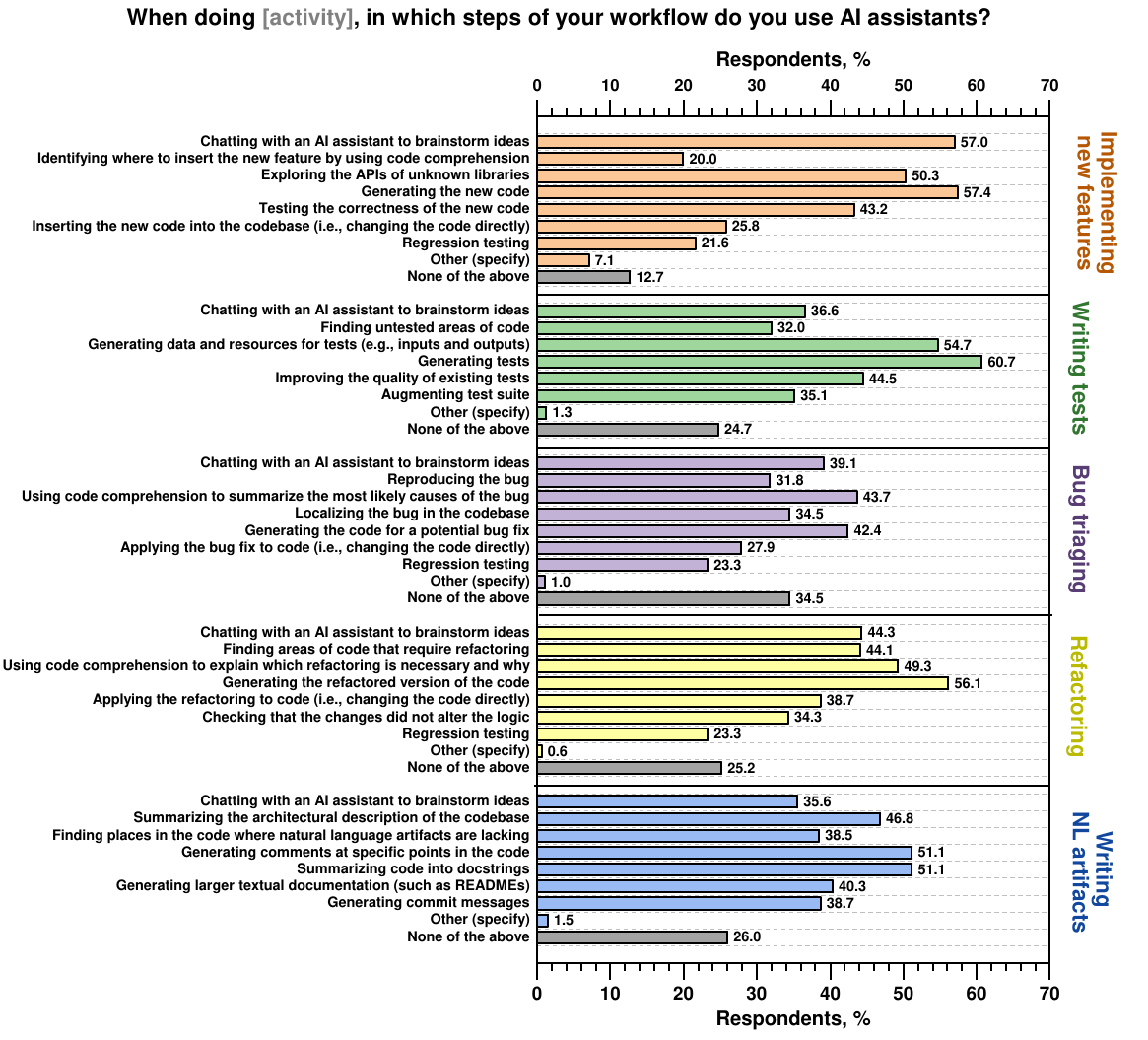}
  \caption{The percentage of respondents who use AI assistants in different stages of different activities.}
  \vspace{1cm}
 \label{fig:stages}
\end{figure}

Now, let us delve into each of the activities and the specific stages within them. The comprehensive results for all stages of all activities can be found in Figure~\ref{fig:stages}.

\textbf{Stages of Implementing new features.} The first thing we can notice about \textit{Implementing new features} is that only 12.7\% of respondents --- the lowest percentage among all activities --- chose the option \textit{None of the above}, meaning that they do not utilize AI tools for any of the stages in this activity. This means that the other 87.3\% use AI in at least one of the stages of this activity, making it the most popular one for AI assistance. Interestingly, in Figure~\ref{fig:dichotomies} we have seen that the smallest percentage of developers wanted to delegate \textit{Implementing new features}, yet in reality the largest percentage of them use AI Assistants for some of the aspects of this activity. This can be explained by the fact that developers assume \textit{full delegation} that they do not want, but actually employ assistants for some stages.

\textit{Chatting with AI assistant to brainstorm ideas}, which is present as a stage in all activities, has the noticeably highest percentage in this one --- 57\% of our respondents chose this option. This indicates the situational importance of conversational functionality for AI assistants.

The other two popular stages of \textit{Implementing new features} are \textit{Generating the new code} (57.4\%) and \textit{Exploring the APIs of unknown libraries} (50.3\%). Interestingly, all the top-3 most popular choices represent different kinds of assistance, indicating the crucial nature of assistants' versatility. 

Among the less popular choices are \textit{Inserting the code into the codebase} (25.8\%) and \textit{Identifying where to insert the new feature by using code comprehension} (20\%), potentially indicating that developers are still more comfortable with making important decisions and changing their code manually.

\textbf{Stages of Writing tests.} Here, we can see that 24.7\% of all participants selected  the option \textit{None of the above}, meaning that three-quarters use AI assistants in this activity.

The most popular stage is \textit{Generating tests}, with as many as 60.7\% of the respondents selecting it. Another popular stage is \textit{Generating data and resources for tests}, which was also selected by more than half of respondents --- 54.7\%. This makes those two stages good cases for current improvement. 

The least selected option is \textit{Finding untested areas of code}, with only 32\% selection rate, once more highlighting that people are less willing to use AI in stages requiring precision. 

\textbf{Stages of Bug triaging.} The first thing that is noticeable from the distribution of respondents for \textit{bug triaging} is the highest percentage of those selecting \textit{None of the above}. As many as 34.5\% of the respondents do not use AI assistants in any way in this activity. 

Another noticeable result is that stages are more equally distributed for bug triaging. The two most popular stages here are \textit{Using code comprehension to summarize the most likely causes of the bug} (43.7\%) and \textit{Generating the code for a potential bug fix} (42.4\%).

Again, one can notice that \textit{Applying the bug fix to code} is among the least popular responses (27.9\%), indicating that here too developers prefer to manually change their code.

\textbf{Stages of Refactoring.} Similarly to \textit{Writing tests}, for \textit{Refactoring}, about a quarter of respondents chose \textit{None of the above}, indicating that three-quarters of them use AI tools at least in some stages.

The most popular response, with 56.1\% of respondents, is \textit{Generating the refactored version of the code}. Another very popular response is \textit{Using code comprehension to explain which specific refactoring is necessary and why}, with 49.3\% of respondents selecting this option. Like the popular code comprehension option in \textit{bug triaging}, this shows the power of the explanatory features of AI assistants. 

Interestingly, though, the choice of \textit{Applying the refactoring to code directly} is more popular (38.7\%) than analogous choices for \textit{Implementing new features} and \textit{Bug triaging}. This might indicate a larger trust in AI when it comes to refactoring or a less critical nature of carrying out refactoring.

Overall, it can be noticed that for \textit{Refactoring}, a lot of options have a high percentage of respondents, indicating the general importance of AI assistance in this activity.

\textbf{Stages of Writing natural-language artifacts.} Similarly to \textit{Writing tests} and \textit{Refactoring}, here also about a quarter of respondents chose \textit{None of the above}, and three-quarters of them use AI assistants in at least one stage.

The two options with the highest percentage of responses --- 51.1\% --- are \textit{Generating comments at specific points in the code} and \textit{Summarizing code into docstrings}. Both of these responses relate to in-code comments, indicating that among code-related natural language artifacts, they are the ones for which people use AI assistants the most. 

The next most common response is \textit{Summarizing the architectural description of the codebase} (46.8\%), again indicating the importance of the explanatory features of AI assistants. It can be noticed, however, that other artifacts---\textit{Larger textual descriptions (such as READMEs)} and \textit{Commit messages}---while being different from regular comments, are also relatively popular to be generated, with 40.3\% and 38.7\% of respondents, respectively. 

\observation{\textbf{Takeaway 3}. 
Generating and summarizing features of AI assistants are usually the most popular. Directly applying the new code to the codebase and searching for places to apply it are less prevalent, although still used by a non-negligible percentage of respondents.}

\subsection{RQ3: Reasons for Not Using Assistants}
\label{sec:results:rq3}

\begin{table*}[]
\caption{The resulting themes in the answers to the question \textit{``What prevents you from using AI assistants in the workflow steps you didn’t select above?''}. * The percentage is calculated from all 1,559 coded responses across all activities.}
\vspace{0.2cm}
\begin{tabular}{ccc}
\begin{adjustbox}{width=1\textwidth,center}
\begin{tabular}{LTc}\hline
\toprule
\multicolumn{1}{c}{\bfseries Theme} & \multicolumn{1}{c}{\bfseries Description} & \multicolumn{1}{c}{\bfseries Responses*} \\
\midrule
\cellcolor{gray!30} Lack of need for AI assistance & \cellcolor{gray!30} Respondents do not try the assistants, express no need or no interest in them, state that the existing non-AI-based tools work well, etc. & \cellcolor{gray!30} 22.5\% \\ 
AI-generated output is inaccurate & The output generated by an AI assistant is inaccurate, the code is incorrect, the model hallucinates, etc.  & 17.7\% \\ 
\cellcolor{gray!30} User's lack of trust and the desire to feel in control & \cellcolor{gray!30} Respondents do not trust the model, consider it to be unreliable, want to feel control over their code, believe that certain things have to be done by people, etc.  & \cellcolor{gray!30} 15.7\% \\ 
Lack of understanding context by AI assistant & The assistant does not understand the context of the task or the underlying reason, cannot analyze the full codebase, cannot understand the requirements or business logic, etc.  & 14.4\% \\ 
\cellcolor{gray!30} User's limited knowledge or understanding of AI assistants and their capabilities & \cellcolor{gray!30} Respondents do not know enough about the AI assistants or how they work, do not know how to do the given task with it, do not know what to ask, etc.  & \cellcolor{gray!30} 10.2\% \\ 
User's desire to perform the task themselves and learn & Respondents want to do the work themselves, love the process, want to understand the code better or to learn, etc.  & 7.3\% \\ 
\cellcolor{gray!30} Time inefficiency & \cellcolor{gray!30} Using AI assistants does not save any time for the respondent, it takes longer to prompt the assistant or to later fix the output than to do the task themsevles, etc.  & \cellcolor{gray!30} 5.5\% \\ 
AI-generated output is not useful & The output generated by an AI assistant is not useful, rudimentary, etc.  & 4.6\% \\ 
\cellcolor{gray!30} Company policies, NDAs, etc. & \cellcolor{gray!30} Respondents are prohibited from using the AI assistants by their companies, NDAs, various policies, etc.  & \cellcolor{gray!30} 3.6\% \\ 
User's negative attitude towards AI & Respondents feel general negative emotions towards AI: hate, not wanting to rely too much, fear of being replaced, etc.  & 3.2\% \\ 
\cellcolor{gray!30} Lack of compliance with non-functional requirements in AI assistant's output & \cellcolor{gray!30} The output generated by an AI assistant does not fit the styling conventions of the project, has poor readability, is too long or too short, is not human-like, etc.  & \cellcolor{gray!30} 3.2\% \\ 
Legal and ethical considerations & Respondents are worried about their privacy, about the copyright of the output, think that using AI is unethical, etc. & 3.1\% \\ 
\cellcolor{gray!30} Challenges with AI integration and usability in development environments & \cellcolor{gray!30} The AI assistant is not integrated well into the IDE, does not have the necessary functionality, has tooling and usability problems, etc. & \cellcolor{gray!30} 3.0\% \\ 
Lack of access & Respondents cannot access AI assistants, they are too expensive, etc.  & 2.6\% \\ 
\cellcolor{gray!30} Security concerns & \cellcolor{gray!30} Respondents are worried about security, the AI assistants providing unsafe code, etc.  & \cellcolor{gray!30} 2.3\% \\ 
Limitations in AI's creativity & AI assistant lacks creativity, cannot provide complex solutions, provides generic code, etc. & 2.2\% \\ 
\cellcolor{gray!30} Workflow disruption & \cellcolor{gray!30} Respondents are not used to using AI assistants, have challenges with adopting it, feel like it breaks their workflow, etc. & \cellcolor{gray!30} 2.1\% \\ 
Limitations of knowledge in the AI model & The model in the AI assistant is outdated, lacks domain knowledge, is not trained well for non-English language, etc. & 1.9\% \\ 
\cellcolor{gray!30} Inefficiency for specific tasks & \cellcolor{gray!30} The AI assistant is incapable of carrying out some specific task, etc. & \cellcolor{gray!30} 1.7\% \\ 
Challenges in communicating user intentions to AI assistant & Respondents find it hard to communicate their intentions to the AI assistant, the assistant cannot follow instructions, requires directly pointing it to the necessary code, etc.  & 1.0\% \\ 

\bottomrule
\end{tabular}
\end{adjustbox}
\end{tabular}
\label{table:themes}
\vspace{-0.4cm}

\end{table*}

Finally, it is critical to understand why developers are hesitant to use AI assistants for some development activities. To do that, we studied their responses to open-ended questions about factors that prevent respondents from using AI assistants. Our thematic analysis resulted in 20 distinct themes. Their full list with detailed description is provided in Table~\ref{table:themes}, with the percentage calculated from all 1,559 coded responses across all five activities. We first discuss the themes in general and then highlight more specific problems for each activity.

\textbf{General themes.} The most prevalent theme is the \textit{Lack of need for AI assistance}, mentioned in 22.5\% of all responses. The respondents highlighted that they did not try assistants for some stages, and sometimes directly mentioned that they do not think the assistance is necessary because they can do the task themselves or they do not perform this task at all. Moreover, sometimes, they mentioned that the existing non-AI-based tools already work well.

The second most popular theme is \textit{AI-generated output is inaccurate}, highlighted by 17.7\% of responses. This group of issues relates to direct inaccuracies of AI-generated output, incorrect code, hallucinations, etc. 

Next is \textit{User's lack of trust and the desire to feel in control}, with 15.7\% of responses. Respondents often mentioned not trusting AI-based tools, the results being unreliable, and their desire to maintain control over the project. This theme is largely prevalent in responses about applying the changes directly to code, confirming our notion from RQ2 that a lot of respondents are not comfortable with this. P380 wrote: \textit{``I wouldn't fully trust the AI to apply a fix and ship it to production without supervision.''}

14.4\% of responses highlighted the theme of \textit{Lack of understanding context by AI assistants}. Sometimes, this relates to more technical and specific things, in particular, the inability of an AI assistant to analyze the full codebase or gain access to third-party code or company-specific artifacts. Some respondents, however, mentioned a broader lack of understanding: understanding the requirements of the task, business logic, etc.

Other, less prevalent themes include \textit{User's desire to perform the task themselves and learn}, \textit{Company policies, NDAs, etc.}, \textit{Workflow disruption}, and others. 

In the following, we highlight specific issues mentioned in different activities. For each activity, we report the three themes, the relative percentage of which in that activity's responses is the highest compared to the overall percentage presented in Table~\ref{table:themes}, together with the reasoning that we collected from the responses.

\textbf{Specifics of Implementing new features.} \textit{AI-generated output being inaccurate} was mentioned in 28.4\% of all coded responses to the question about implementing new features. Here, incorrect code and hallucinations are especially critical, therefore, people encountering them are cautious in using AI for this activity. Some respondents mentioned that the AI assistant's output requires careful editing and that the errors are sometimes subtle and difficult to fix. P768 wrote: \textit{``In my experience, AI's outputs very often contain subtle errors, which is why I have become very cautious.''}

Another aspect important for new features is \textit{Time inefficiency}, raised in 11.3\% of answers for this activity. Some respondents mentioned that correctly prompting the AI assistant takes longer than writing the code themselves, and some mentioned that fixing the generated code takes longer. 

\textit{Workflow disruption} was mentioned in 3.5\% of responses regarding \textit{Implementing new features}, with respondents not being used to utilizing assistants or mentioning that they interrupt their flow. P245 explained this in detail: \textit{``...In my head I am several steps ahead of the code which is currently being entered into the file, and the AI assistants force me to backtrack constantly to analyze whether what they wrote was what I actually meant to write.''}

\textbf{Specifics of Writing tests.} For \textit{Writing tests}, as many as 32.1\% responses mentioned \textit{Lack of need for AI assistance}. Some respondents do not do testing at all, and some mentioned that the existing tools work well for finding untested areas of code. P98 wrote: \textit{``There are multiple extensions for both Rider and Visual Studio as well as commandline tools that show code coverage, so there is no need to use AI for that.''}

Also, 7.8\% of responses about tests highlighted that \textit{AI-generated output is not useful}, and 2.9\% of responses mentioned \textit{Limitations in AI's creativity}. Specifically, respondents mentioned that the generated tests are sometimes useless, rudimentary, and that the assistants cannot provide complex solutions. P525 wrote: \textit{``The generated code is not relevant most of the time, ... It is suitable for elementary tests only.''}

\textbf{Specifics of Bug triaging.} 18.5\% of responses related to \textit{Bug triaging} mentioned \textit{Lack of understanding context by AI assistant}. Respondents mentioned that finding and fixing non-trivial bugs requires a large context, both in terms of code and in terms of a more high-level, conceptual understanding of the program, which the AI assistants lack. P253 wrote: \textit{``A bug can come from a subpar execution of an idea or because a previously unknown behavior of the library/programming language/api. Investigating and fixing such bugs requires understanding a great deal of information...''}

Also, 3.1\% of responses in \textit{Bug triaging} mention \textit{Inefficiency for specific tasks}, specifically mentioning that AI assistants are not capable of finding bugs or fixing them. 

Finally, \textit{Bug triaging} is the main source for \textit{``Challenges in communicating user intentions to AI assistant''}, with 2.5\% of responses mentioning it. P128 wrote: \textit{``AI assistants can't understand all the details why there might be a bug because you can't describe it (otherwise you would know the problem).''}

\textbf{Specifics of Refactoring.} For \textit{Refactoring}, 21.9\% of responses mentioned \textit{User's lack of trust and the desire to feel in control}. This is important for refactoring because it is crucial for the refactored code to not alter the logic of the original code. To this point, P60 wrote: \textit{``I prefer to make sure myself that the new refactored code does not change the logic.''}

Also, as many as 10.4\% of responses mention \textit{User's desire to perform the task themselves and learn}. Some respondents use refactoring as an opportunity to understand the code better and some just enjoy this activity overall. P558 mentioned this in their response: \textit{``Refactoring is one of the most enjoyable aspects of coding.  It not only improves the readability and reusability of code but helps better understand it.''}

4\% of responses in this activity also mention \textit{Lack of compliance with non-functional requirements in AI assistant's output}, highlighting that the code refactored by the AI assistant may lose readability and not adhere to the project's preferences. P47 wrote strongly: \textit{``...Auto-generated code is usually total garbage when it comes to structure and formatting.''}

\textbf{Specifics of Writing natural-language artifacts.} 29\% of responses about natural-language artifacts contained the theme of \textit{Lack of need for AI assistance}. Interestingly, Section~\ref{sec:results:rq2} and Figure~\ref{fig:dichotomies} indicated that developers strongly want to delegate this activity. However, a lot of respondents said that they did not try AI assistants for certain artifacts, especially larger ones like READMEs. Moreover, some of them indicated that the existing tools work fine for them.

A crucial theme for natural artifacts is \textit{Lack of compliance with non-functional requirements in AI assistant's output} (9.5\% responses). Respondents mentioned that the text provided by the assistant does not sound human-like, may be bloated, verbose, not fitting the conventions of the projects, or overall not useful to the human reader. P273 succinctly put it like this: \textit{``Language produced by LLMs is often unnatural and does not match the required tone and clarity.''}

Finally, 2.7\% of responses about this activity mention \textit{Workflow disruption}, with P356 writing simply and directly: \textit{``I'm not used to it yet''}.

\textbf{Overcoming the drawbacks.} At the very end of each block, we asked the participants whether they would consider using the AI assistant for the stages they did not select if it did not have the described drawbacks. The detailed figure can be found in our supplementary materials~\cite{artifacts}. While \textit{Writing tests} and \textit{Writing natural-language artifacts} are still the most likely to be delegated (71\% and 72\%, respectively), all the other three activities are also over 60\% positive, indicating that in principle, the respondents are open to using AI assistants if the shortcomings are addressed.

\observation{\textbf{Takeaway 4}. The top reasons why developers are not using AI assistants are the lack of need, AI-generated output being inaccurate, lack of trust, and the lack of understanding of context by the assistant.}
\section{Discussion \& Implications}
\label{sec:discussion}

The results of our survey can provide a user-centered approach to focus the future development of AI assistants. The homogeneous comparison of different activities and different stages within them can help us select the most promising and the most problematic aspects of assistance usage. 

\subsection{What Needs Our Focus Now}
\label{sec:discussion:now}

The results of our study support the existing notion about the current state of affairs of using AI assistants in software development as favorable and rapidly developing~\cite{liang2024large, Ziegler2022Productivity}. We see that people adopted the technology in their workflows and see the code provided by AI as usable and fairly accurate. However, in line with previous studies~\cite{287298, asare2023github, 9833571}, we see that people are cautious regarding AI-generated code being insecure. We perceive this serious drawback as an opportunity for the community to research and develop solutions to provide users with more safe and secure technology.

Our high-level comparison of activities (Section~\ref{sec:results:rq2}) indicates that in terms of activities, \textit{writing tests} and \textit{writing natural language artifacts} are good candidates for improvement. The comparison of responses to different questions allows us to conclude that respondents find these tasks to be somewhat unpleasant and express their desire to have them automated. Our detailed subsequent questions about individual stages  highlight how specifically respondents use AI assistants within these activities: \textit{generating tests}, \textit{generating data and resources for tests (\textit{e.g.} inputs and outputs)}, \textit{summarizing code into docstring}, \textit{generating comments at specific points in the code}. While some of these tasks are well-established, others are less studied --- for example, using AI assistants for generating separate resources for tests requires more attention from the research community. Combining the established collaboration between developers and AI on these tasks with the willingness to delegate activities to AI, they warrant attention from the community for possible innovations in the respective fields.

In the other studied activities, we can see some common patterns that are also important to take into account for improving the support of AI tools. In this regard, among the most promising steps in \textit{Implementing new features}, \textit{Bug triaging} and \textit{Refactoring} are \textbf{(a)} the generation of the new necessary code and \textbf{(b)} the summarization of code for a better understanding of the context of the change. These two activities align with the grounded theory proposed by Barke et al.~\cite{Barke2023Grounded}, which describes two developers' interaction modes with AI assistants --- acceleration (supported by code generation) and exploration (supported by code summarization). Our detailed results for individual stages can highlight some more specific candidates for focus --- \textit{Using code comprehension to summarize the most likely causes of the bug}, \textit{Generating the refactored version of the code}, and others. We suggest that studies of possible improvements in those directions would be important for developers and impactful for the field.

Finally, while the option to chat with the assistant is especially crucial when \textit{implementing new features}, it is present in all the other activities as well. This indicates the importance of not only the ``technical'' aspects of the models (\textit{e.g.}, generating quality code and summarizing the code correctly) but also UI and UX-related aspects, as mentioned by Zhou et al.~\cite{zhou2023concerns}, specifically conversational functionality of AI assistants as highlighted in works of Ross et al.\cite{ross2023programmer} and Ziegler et al.~\cite{Ziegler2022Productivity}. 

These results can be formulated into the following general implications:
\begin{itemize}
    \item \textbf{For researchers}. To bring the most immediate value in the areas where developers already highlight their needs, one can focus on the innovation around the systems that ensure usage of AI assistants in writing tests, writing natural language artifacts, code generation, and code summarization.
    \item \textbf{For tool builders}. The specific features that developers are already actively using --- generating tests, summarizing code into docstrings, etc. --- are crucial for better integration into the existing tools, such as IDEs. Some of the used features might not yet be directly available in the workflow, such as generating data for the tests, and it is worth experimenting with them.
\end{itemize}

\subsection{What Needs to Change in the Future}
\label{sec:discussion:future}

Results presented in Section~\ref{sec:results:rq3} indicate a way towards greatly improving the AI experience for users. While many different reasons for not using assistants can be seen in Table~\ref{table:themes}, still the majority of the respondents said that they are likely or very likely to use AI assistants for all activities if the shortcomings were overcome. The large-scale nature of our survey provides both technical and more fundamental reasons, which makes it difficult to address them equally. For example, \textit{Lack of need for AI assistance} is a more general thing that gradually changes over time. However, the wide array of answers from our respondents, referring to multiple different activities and stages within them, allows us to propose a multi-faceted look at the future research to tackle the existing shortcomings.
We see the possible space of innovation threefold: system, integration, and education.

\textbf{One direction is the improvement of base systems of AI assistants.} Addressing \textit{AI-generated output being inaccurate} and the \textit{Lack of compliance with non-functional requirements} is crucial, with 17.7\% and 3.2\% of all responses mentioning these issues, respectively, which is in line with the study of Liang et al.~\cite{liang2024large}. Previous research shows that developers spend almost 50\% of coding time in interaction with the LLM, with 35\% dedicated to double-checking suggestions, since the generated code is limited in meeting both functional and non-functional requirements~\cite{mozannar2023reading}. Therefore, resolving accuracy issues would also affect the \textit{Time inefficiency} of AI tooling usage, mentioned in 5.5\% of all responses. 

It is also important to resolve \textit{Lack of understanding context by AI assistants}. This issue was raised in 14.4\% of all responses and represents one of the most specific and concrete reasons for not using AI assistants. One part of it is the inability of the models to consider the entire codebase of the project. Another aspect of this issue is the assistant's lack of access to other sources of information, such as project-specific documentation, internal knowledge bases, issue trackers, etc. In this line lies also the \textit{Limitations of knowledge in the AI model} and consequent \textit{Inefficiency for specific tasks}, mentioned in 1.9\% and 1.7\% of participants' answers, respectively. In this regard, a pivotal direction for future research is finding ways to compress large amounts of information and project-level context into the model~\cite{lozhkov2024starcoder, crosscodeeval}.

Moreover, \textit{Legal and ethical considerations} and \textit{Security concerns} regarding the use of AI assistance, which came up in 3.1\% and 2.3\% of responses, respectively, could be addressed via system changes~\cite{licenses-lawsuit}. It is promising to investigate local models that can work on the user devices and not send the data over the internet (\textit{e.g.}, federated learning)~\cite{li2020review}. This is also important for \textit{Company policies, NDAs, etc.}, the issues about which were raised in 3.6\% of responses.

We believe that if accuracy, understanding, and compliance with other quality-related requirements were fixed, people would also probably gain more trust receiving more useful and meaningful assistance, resolving the other popular reasons not to use AI---\textit{Lack of trust} and \textit{AI-generated output not being useful}---raised in 15.7\% and 4.6\% of responses respectively and widely discussed in the community~\cite{liang2024large, eudigitalstrategy, amoozadeh2023trust}. 

\textbf{Another possible direction for research and innovation is integrating AI systems into developers' workflow.} We believe that such issues as \textit{Challenges with AI integration and usability in development environments} (raised in 3\% of all responses) and \textit{Workflow disruption} (raised in 2.1\% of responses) should be addressed by companies who provide AI assistants for the market, considering their hands-on access to the systems and their users for more quick and meaningful research and innovation~\cite{vaithilingam2023towards, mozannar2023suggestion}.

\textbf{Moreover, it is important to educate and inform users and potential users about the capabilities and limitations of AI assistants.} Mentioned in 10.2\% of responses, \textit{User’s limited knowledge or understanding of AI assistants and their capabilities}, and in 1\% of responses --- \textit{Challenges in communicating user intentions to AI assistant}, these drawbacks may be addressed by the wider and specific knowledge-sharing activities in the community and beyond it~\cite{feng2024coprompt, chi2024workshop}. Moreover, with the evolution of AI-supported educational activities~\cite{chen2024learning, Robe2022Designing,Jayagopal2022Exploring}, the community of AI-aware developers will expand greatly in the near future.

These results can be formulated into the following general implications:
\begin{itemize}
    \item \textbf{For researchers}. With the recent advances in developing AI assistants, a lot of different problems remain, all of which require long-term research. In particular, it is important to focus not only on the systems themselves --- addressing common issues like lack of quality, understanding larger context, and ensuring legal and security compliance --- but also on the human element. Research should align the system with human needs, understanding the reasons behind certain challenges to effectively resolve them and create more user-centered solutions.
    \item \textbf{For tool builders}. A good AI model in itself is not enough to bring value to the user. It is crucial to develop novel ways of integrating AI assistants into the developers' workflow and invest in better UI/UX. With industrial experience in these kinds of improvements, this can be handled by the development teams of tools such as IDEs.
    \item \textbf{For developers}. While the tools should improve their education of users on AI assistants, it is also on developers to inform themselves. Better knowledge of various features and experimenting with new ones can vastly improve the experience, as noted by some responses to our questions that highlighted the desire of the participants to try out new things that they picked up in our survey.
\end{itemize}

In essence, the findings we gathered offer valuable direction for both short-term research goals and long-term strategies. While developers have voiced a variety of concerns, we are confident that with the concerted efforts of the research community, these issues can be effectively addressed, ultimately leading to a more robust and beneficial user experience.

\section{Threats to Validity}
\label{sec:ttv}

The large-scale and general nature of our study introduces several important threats to the validity of our study.

\textbf{Generalizability}. Our results are based on a specific sample of people and might not generalize. In particular, even though the mailing list curated by JetBrains includes users of different IDEs and AI assistants, it can introduce some bias into the results. At the same time, our sample of 481 people is large for studies in our field~\cite{Medeiros2018Discipline}, and it is diverse in terms of programming languages and types of developed software.

\textbf{Thematic analysis}. It is possible that our thematic analysis resulted in missing some themes, which could influence our analysis. To combat this, we carefully followed established practices~\cite{fereday2006demonstrating}: two authors came up with codes and themes independently and then reached an agreement in the discussion, and we also removed from the analysis responses that did not clearly formulate their issues.
\section{Conclusion}
\label{sec:conclusion}

In this paper, we set out to study specific ways in which developers use AI assistants in different stages of the software development life-cycle, as well as the reasons why they do not use them. Our results show that respondents use AI assistants in various activities and various stages within them, but unequally. Developers indicated \textit{Writing tests} and \textit{Writing natural language artifacts} as the least enjoyable activities that they would want to delegate to an AI assistant. Within different activities, developers tend to use assistants for generation and summarization, and employ them for finding places in the code and applying code more rarely. In terms of the reasons for not using AI assistants, our thematic analysis revealed 20 diverse issues, with the main ones being \textit{Lack of need for AI assistance}, \textit{AI-generated output being inaccurate}, \textit{Lack of trust}, and \textit{Lack of understanding of context by AI assistant}. We also highlight issues raised by the participants in regard to individual activities, which can inform further research. The survey and its full results can be found in supplementary materials~\cite{artifacts}.

We believe our work is especially needed right now, when AI assistance is being developed rapidly, since our comprehensive and specific results can be used to guide further research and implementation.

\section*{Acknowledgements}

We would like to extend our sincere gratitude to the Surveys Team and the Research Core Team of JetBrains for their invaluable support and assistance throughout this study. Their expertise and dedication were crucial in facilitating the research process and ensuring its success. We greatly appreciate their contributions and collaboration.
 \bibliographystyle{elsarticle-num} 
 \bibliography{bib.bib}






\end{document}